\documentclass[12pt]{iopart}
\usepackage{graphicx,amssymb,subfigure,float}

\newcommand{\rv}{\bi{r}}
\newcommand{\pv}{\bi{p}}

\begin{document}
\title{Transition to hydrodynamics in colliding fermion clouds}
\author{F.  Toschi$^{1,2}$, P. Capuzzi$^{3}$, S. Succi$^{1,3}$,
P. Vignolo$^3$, and M. P. Tosi$^3$}

\address{$^1$ Istituto per le Applicazioni del Calcolo, CNR, Viale del
Policlinico 137, I-00161 Roma, Italy} \address{$^2$ INFM, Unit\`a di
``Tor Vergata'', Via della Ricerca Scientifica 1, I-00133 Roma, Italy}
\address{$^3$ NEST-INFM and Classe di Scienze, Scuola Normale Superiore,
I-56126 Pisa, Italy }

\ead{toschi@iac.cnr.it}

\begin{abstract}
  We study the transition from the collisionless to the hydrodynamic
  regime in a two-component spin-polarized mixture of $^{40}$K atoms
  by exciting its dipolar oscillation modes inside harmonic traps.
  The time evolution of the mixture is described by the Vlasov-Landau
  equations and numerically solved with a fully three-dimensional
  concurrent code. We observe a master/slave behaviour of the
  oscillation frequencies depending on the dipolar mode that is
  excited. Regardless of the initial conditions, the transition to
  hydrodynamics is found to shift to lower values of the collision
  rate as temperature decreases.
\end{abstract}
\submitto{\jpb}


\maketitle

\section{Introduction}

Several experiments on cooling two spin-polarized states of a Fermi
gas have proven that this setup is an important tool for investigating
the collisional properties of such a quantum system
\cite{Gensemer2001a}.  The JILA group has recently performed
experiments in which the collisionality of a $^{40}$K mixture is tuned
by either varying the atomic density \cite{Gensemer2001a,DeMarco2002a}
or the off-resonant value of the inter-species scattering length
\cite{Loftus2002a,Regal2003b}.  Also the temperature of the gas has
been used to drive its collisionality and the importance of Pauli
blocking at very low temperatures has thereby been 
demonstrated~\cite{Gensemer2001a}.  The role of
collisions has also been investigated in dipolar collective modes of a
boson-fermion mixture~\cite{Ferlaino2003a} at temperatures in both
the quantum and classical regimes.  The behaviour of the collision
rate inferred from these experiments plays a crucial role in the
development of strategies to improve the evaporative cooling
techniques.  In addition, the collisionality also affects other
observables such as the damping and diffusion processes
in the quantum mixture.  In
Ref.~\cite{Regal2003b} an anisotropic expansion of a cigar-shaped
two-component Fermi gas has been observed. In contrast to the bosonic
case, the anisotropy of the expansion at very low temperatures cannot
be taken as a clear signature of fermionic superfluidity
\cite{Ketterle2003a} and could be due to purely collisional
effects.

While ergodic approximations on semiclassical Boltzmann equations have
been used to describe the evaporative cooling
process~\cite{Geist2002a}, a microscopic understanding of the
transition towards hydrodynamics \cite{Gensemer2001a} or of the
anisotropic expansion in the case of attractive
interactions~\cite{Regal2003a} needs a complete numerical study of the
dynamics of the quantum gas in position and momentum space
\cite{Toschi2003a}.  In this paper we describe the strategy that we
have used to solve numerically the Vlasov-Landau Equations (VLE) for
an ultracold two-component fermion gas with a fully three-dimensional
(3D) algorithm.  The two fermionic fluids are treated by a
particle-dynamics approach~\cite{Bird1994a}, accounting for mean-field
interactions and for collisions between the two species.  Since the
Pauli principle causes a saturation of phase space at very low
temperature, we have developed a {\it locally adaptive
importance-sampling} technique which allows us to select the colliding
particles in a way which is by several orders of magnitude faster than
in standard Monte Carlo techniques.  Furthermore, the control of the
occupancy of the unitary cells of phase space and a suitable choice of
{\it computational parameters} permit us to avoid Pauli
inconsistencies during the dynamics.

The two fermionic gases are confined inside different harmonic traps.
Regardless of the initial perturbation exciting the dipolar modes, we
observe that as temperature is lowered, even through most collisions become
forbidden classically and by the Pauli principle, the few collisions
that occur still suffice to drive the mixture from the collisionless
to the hydrodynamic regime.  This effect has already been predicted in
our previous study~\cite{Toschi2003a}, where the angular degree of
freedom was taken into account {\it via} an effective weight in a 2D
numerical code. The limitations of the 2D approach are illustrated in
the present work by comparing the oscillation frequencies and the
damping rate of the dipolar oscillations evaluated within the 2D
approach with those obtained by means of the present 3D scheme. We
also find that the choice of the initial state modifies the spectra of
the dipolar modes. In particular, if the symmetry of the initial
configuration is broken and one species is initially at rest, kicking
by the other species will induce oscillations at the characteristic
frequency of the latter.

The paper is organized as follows. In Sec.~\ref{sec2} we introduce the
physical model and describe in detail the numerical method developed
to solve it.  In Sec.~\ref{sec3} we study the dipolar oscillations in
various collisional regimes and compare the results with those
obtained in our previous work.  Concluding remarks are given in
Sec.~\ref{sec4}.

\section{\label{sec2}The model and its solution in 3D}

The two fermionic components in external potentials $V_{\rm
ext}^{(j)}(\rv)$ are described by the distribution functions
$f^{(j)}(\rv,\pv,t)$ with $j=1$ or 2.  They obey the kinetic
equations
\begin{equation}
\partial_t f^{(j)}+\frac{\pv}{m}\cdot{\bi{\nabla}}_{\rv} 
f^{(j)}-
{\mathbf{\nabla}}_{{\rv}} U^{(j)}\cdot {\bi{\nabla}}_{\pv}f^{(j)}=
C_{12}[f^{(j)}]
\label{vlasov}
\end{equation}
where the mean-field effective potential is $U^{(j)}(\rv,t)\equiv
V_{\rm ext}^{(j)} (\rv)+g n^{(\bar{j})}(\rv,t)$ with $\bar{j}$
denoting the species different from $j$.  Here we have set $\hbar=1$,
$g=2\pi a/m_r$ with $a$ being the $s$-wave scattering length between
two atoms of different species and $m_r$ the reduced mass, and
$n^{(j)}(\rv,t)$ is the density obtained by integrating 
$f^{(j)}(\rv,\pv,t)$ over momentum.

Collisions between atoms of the same spin can be neglected at low
temperature, so that in Eq.\ (\ref{vlasov}) the term $C_{12}$ involves
only collisions between particles that are polarized in two different Zeeman
states. We have:
\begin{equation}
C_{12}[f^{(j)}]\equiv 
g^2\,\frac{2(2\pi)^4}{V^3}\!
\sum_{\pv_2,\pv_3,\pv_4} 
\Delta_{\pv}\Delta_{\varepsilon}
\,[{\bar{f}}^{(j)}\bar{f}_2^{(\bar{j})}f_3^{(j)}
f_4^{(\bar{j})}-
f^{(j)}f^{(\bar{j})}_2\bar{f}_3^{(j)}
\bar{f}_4^{(\bar{j})}]
\label{integralcoll}
\end{equation}
with $f^{(j)}\equiv f^{(j)}(\rv, \pv,t)$, $\bar{f}^{(j)} \equiv
1-f^{(j)}$, $f_i^{(j)}\equiv f^{(j)}(\rv,\pv_i,t)$, $\bar{f}_i^{(j)}
\equiv 1-f_i^{(j)}$.  $V$ is the volume occupied by the gas and the
factors $\Delta_{\pv}$ and $\Delta_{\varepsilon}$ are the usual delta
functions accounting for conservation of momentum and energy, with the
energies given by $p_i^2/2m_j+U^{(j)}(\rv,t)$.

The equilibrium state of the mixture is given by the stationary
solution of Eq.\ (\ref{vlasov}), {\it i.e.} by the local
Fermi-Dirac distributions
\begin{equation}
f^{(j)}(\rv,\pv) = \left(e^{\beta\left(p^2/2m_j + U^{(j)}(\rv)
    -\mu_j\right)} +1\right)^{-1}
\end{equation}
at given temperature $T = 1/k_B\beta$, 
where $\mu_j$ is the chemical potential ensuring the normalization
condition $\int f^{(j)}(\rv,\pv) d^3r\,d^3p/h^3 = N_j$. The particle
densities entering $U^{(j)}$ are to be determined self-consistently by
integration over momenta. The details of this calculation have been
previously given in Ref.\ \cite{Amoruso1999a}.

\subsection{Numerical method} 
The numerical solution of the coupled VLE in Eq.\ (\ref{vlasov}) is
carried out inside a finite box of size $L_x\times L_y\times L_z$ and
discretized with a spacing $\Delta x_j$. We describe each
species in the box by means of a set of $N_c$ computational particles (cp) 
at phase-space points $\{\rv_i,\pv_i\}$ which are spread according to the
distribution $f$. In this way each fermion is represented by
$N_q=N_c/N_j>1$ ``quarks''.

The initial selection of the computational particles is made by direct
Monte Carlo sampling of the equilibrium distributions.  
%
Moreover, since collisions are very sensitive to the statistical
nature of the colliding particles, we enforce the Pauli principle in
each phase-space volume of size $h^3$ by resampling the computational
particles that would otherwise exceed the allowed number of quarks
$N_q$ in that cell.
This
further control is needed since we are dealing with a finite-size
sample of Fermi-Dirac distributions.

The solution of Eqs.\ (\ref{vlasov}) proceeds in two steps: (i) the
Lagrangian evolution of each distribution in phase space due to 
mean-field and external forces, and (ii) the occurrence of binary
collisions among particles of the two species {\em \`a la} Boltzmann.  Once the
initial state is prepared, the first stage of the evolution is carried
out by means of a second-order symplectic integrator
\cite{Nettesheim1996a,Jackson2002b}, and we update the
position and velocity of each particle $i$ according to the scheme
\begin{equation}
\left\{
\begin{array}{lcl}
\tilde{\bi{r}}_i &=& \bi{r}_i(t) + \frac{1}{2}\,\Delta t\,\bi{v}_i(t),
\nonumber \\
\bi{v}_i(t+\Delta t) &=& \bi{v}_i(t)  +m^{-1} \, \Delta t\, 
\bi{F}(\tilde{\rv}_i) \\
\rv_i(t+\Delta\,t) &=& \tilde{\rv_i} + \frac{1}{2}\,\Delta t
\,\bi{v}_i(t+\Delta t)\\
\end{array}\right. .
\label{symplec}
\end{equation}
Here $\bi{F}(\rv)$ is the mean-field force $\bi{F}(\rv)=-\nabla
U^{(j)}(\rv)$ with $j= 1$ or 2.  To compute the mean-field forces we
first calculate the average density at a given cell by counting the
number of particles inside the cell. The density is then numerically
differentiated to obtain the forces at the grid points. The forces so
calculated are finally linearly interpolated to the positions of the
particles.

It is worthwhile remembering that Liouvillean evolutions are
incompressible, so that a distribution function satisfying the Pauli
principle at a given time will satisfy the same constraint
afterwards. This is an exact property whose {\it numerical}
counterpart is checked during the course of the simulation.

The second step of the solution involves binary collisions. At
variance from the Lagrangian evolution these are tracked on a coarser
mesh of spacing of the order of the de~Broglie wavelength (a ``Pauli
cell''). In this way we introduce a temperature-dependent scale which
imposes a length scale on which the particles are effectively
indistinguishable. This can be seen as an extension of the standard
Boltzmann approach. Furthermore, collisions in different Pauli cells can be
handled independently due to the locality of
Eq.~(\ref{integralcoll}).
 
Our method allows us to count the number of collisions that occur
step-by-step both classically and quantum-mechanically. In each Pauli
cell we add up the classical probability of collisions between every
pair of particles and store the pairs whose probability is greater
than a {\em locally adapted} threshold.  This reduces the running time
and the storage needs of the code by several orders of magnitude, and
at the same time the threshold is adjusted in order to guarantee a
correct supply of pairs at each time step. The quantum features of
the mixture are included in the last step of the collision algorithm
where the Pauli suppression in Eq.\ (\ref{integralcoll}) is
applied. For every classically colliding pair we calculate the
blocking terms $(1-f^{(j)})$ by analyzing the occupancy of the
momentum states in the Pauli cell. We have explicitly verified that
this procedure yields the correct Pauli-suppressed fraction of
collisions for a weakly interacting gas at least down to $T\sim
0.2\,T_F$ \cite{Succi2003a}.

We turn now to the discussion of the choice of computational
parameters which allows us a reliable implementation of the Lagrangian
evolution. During the dynamics two unphysical processes may occur: (i)
the presence of an excess of quarks in saturated phase-space cells,
and (ii) a broadening of the atomic cloud inside the trap. No cell of
volume $h^3$ contains more than $N_q$ quarks for each species if the
initial state has been accurately prepared.  In this case, the
inconsistent location of quarks takes place {\em after} we move the
particles.  Indeed, both problems (i) and (ii) are related to the
calculation of the accelerations induced by mean-field forces. As
already mentioned, these forces are constructed first on the
simulation grid and then interpolated to the actual position of each
particle. Therefore, an accurate description requires small spacings
$\Delta x_j$ and a large number of computational particles per cell.

Aiming at understanding the impact on the results of the different
choices for the values of $N_q$ and $L_j/\Delta x_j$ we have studied
the axial motion of two interacting Fermi clouds for several values of
these parameters.  Hereafter we shall consider magnetically trapped
$^{40}$K atoms that are equally shared in two different Zeeman states
($m_F=9/2$ and $m_F=7/2$) and confined in isotropic harmonic traps
with slightly different frequencies ($\omega_{9/2} = 2\pi \times
19.8\,{\rm s}^{-1}$ and $\omega_{7/2} = 2\pi \times 17.46\,{\rm
s}^{-1}$). For this test case we take the total number of particles
$N=200$, the mutual scattering length $a=3\times10^4$ Bohr radii (a
strongly collisional regime), and $T=0.3\,T_F$. Both clouds are
suddenly displaced from their equilibrium positions along the axial
direction and start oscillating. We then analyze the evolution of the
mixture for a total time $t_{fin}\simeq 50\,\omega_{9/2}^{-1}$
($\sim8$ complete oscillations) and from the value of the
centre-of-mass position $z_j(t)$ we extract the oscillation
frequencies $\omega_j$ of the two components and the average damping
rate $\bar{\gamma}$. The details of this procedure will be given in
Sec.~\ref{sec3}. In addition we also compute the axial spreads of the
density profiles $D_j=\langle \Delta z_j\rangle_{t_{fin}}/\langle
\Delta z_j\rangle_{t_{0}}$ and the quantum collision rate $\Gamma_q$
by direct counting of the total collisions per particle, and monitor
the number ${\cal N}$ of computational particles located in forbidden
phase-space cells. The latter shall be called Pauli {\em
inconsistencies}. For the system parameters here considered, we find
that the oscillation frequencies do not vary considerably as $N_q$ and
$L_j/\Delta x_j$ are varied.  The dependence of the other observables
is summarized in Table~\ref{tavola}.

\begin{table}
\caption{\label{tavola}The maximum relative value $\tilde{\cal
N}=({\cal N}/N_c)_{\rm max}$ of particles in excess per time step, the
maximum spread $D=\max\{D_{9/2}, D_{7/2}\}$ of the clouds
in the axial
direction, the collision rate $\Gamma_q$ (in units of 1/s), and the
average damping rate $\bar{\gamma}$ (in units of 1/s) as functions of
the number of quarks $N_q$ and of grid points $L_j/\Delta x_j$. The data
correspond to the two-component Fermi gas at $T=0.3T_F$ with a total
number of particles $N=200$ and an inter-species scattering length
$a=3\times 10^4a_0$ ($a_0$ being the Bohr radius).}
\begin{indented}
\item[]\begin{tabular}{@{}p{4cm}cccc}
\br
 & $\tilde{\mathcal{N}}$ & $D$ & $\Gamma_q$ (1/s) & $\bar\gamma$ (1/s) \\
\mr
$L_j/\Delta x_j=100$ & & & &\\[2pt]
$N_q = 20$  & $4\times10^{-3}$ &  1.40 & 21.10 & 1.26 \\
$N_q = 50$ & $7\times10^{-3}$ & 1.20 & 23.05 & 1.33 \\
$N_q = 200$ & $6\times10^{-2}$ & 1.00 & 21.60 & 1.07 \\[5pt] 
$L_j/\Delta x_j=200$ & & & & \\
$N_q = 200$ & 0 & 1.40 & 21.80 & 1.50 \\
$N_q = 400$ & 0 & 1.25 & 23.50 & 1.30 \\
$N_q = 1600$ & 0 & 1.00 & 29.10 & 1.06 \\
\br
\end{tabular}
\end{indented}
\end{table}

The inconsistencies disappear by increasing the number of grid points,
while a large number of computational particles per cell is required
to prevent statistical noise from producing spurious diffusion of the
particles. The average damping rate $\bar\gamma$ is sensitive to the
spreading of the clouds, but seems to be less affected by the presence
of few inconsistencies. On the other hand the collisional rate
$\Gamma_q$ depends more strongly on both factors.

In summary, this test case has given us enough information 
to select appropriate parameters for reliable
calculations. In the following numerical experiments we shall choose
the most suitable values of $N_q$ and $L_j/\Delta x_j$ in order to
ensure $D_j\simeq 1$ and a negligible ${\cal N}$, while keeping 
the computing time within reasonable limits.

\section{\label{sec3}Dipolar oscillations}

The quantum features of the mixture are most conveniently displayed by
its collective modes, which in turn can be investigated by analyzing
the dynamical response of the system under distortions of the
equilibrium state. This can be done quite straightforwardly in actual
experiments where the cloverleaf setup is used \cite{Gensemer2001a,%
DeMarco2002a, Loftus2002a, Regal2003b}: a bias magnetic field
displaces the centre of the trap, while a Stern-Gerlach separation of
the two Zeeman states could be obtained by adding a magnetic field
gradient \cite{Gensemer2001a,Stamper-Kurn1998a}.

In the following we analyze the dipolar modes of the
mixture at varying scattering length, as can be done {\it e.g.} by
exploiting Feshbach resonances \cite{Loftus2002a}, and
focus on two specific types of initial conditions: one in which both
components start oscillating with the same amplitude 
({\em in-phase oscillations})
and the other in which only the $m_F=9/2$ species is initially
displaced ({\em kicked oscillations}).

\subsection{In-phase oscillations}

\begin{figure}

\subfigure[]{\includegraphics[width=0.52\linewidth]{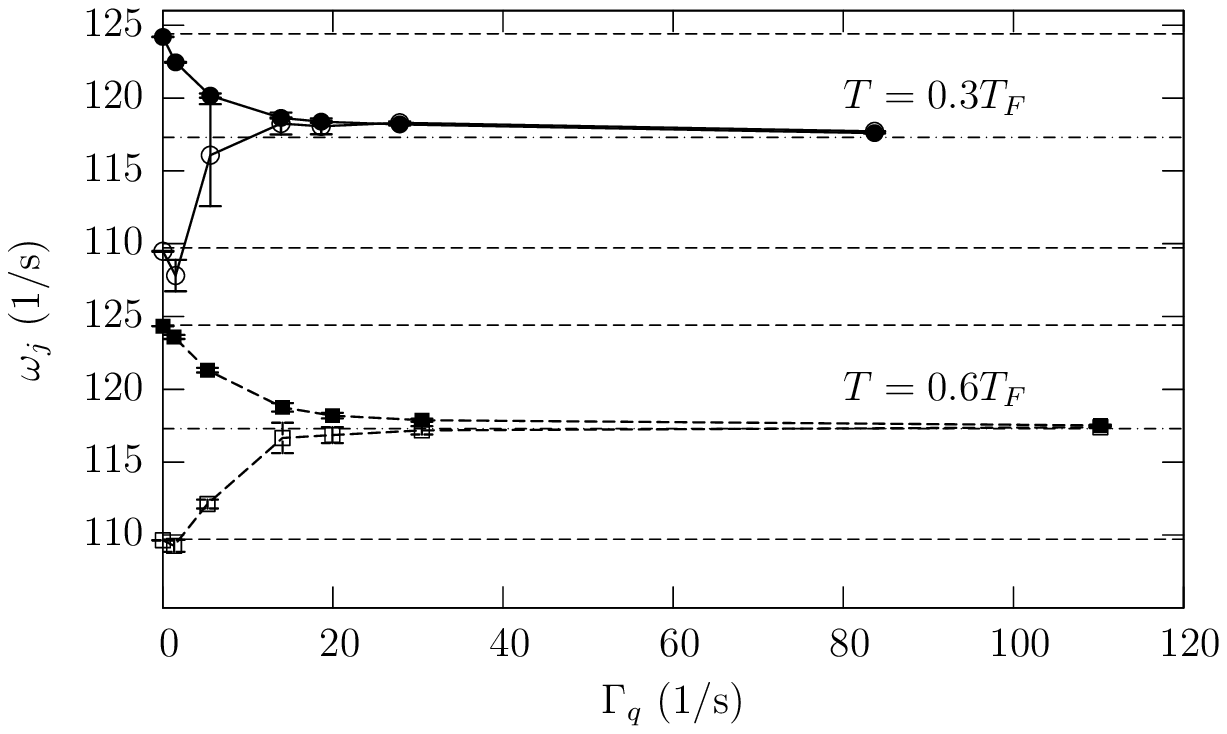}}
\subfigure[]{\includegraphics[width=0.52\linewidth]{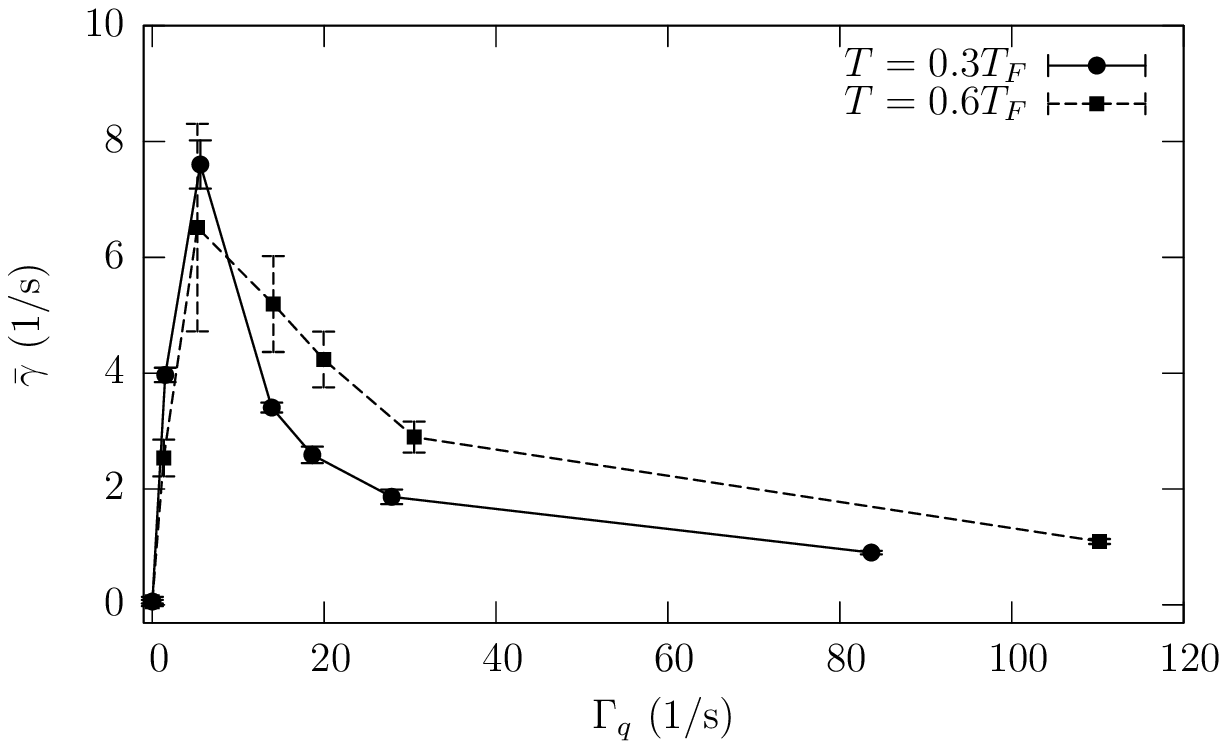}}
\caption{Transition from the collisionless to the collisional regime
  in a $^{40}$K mixture with $N_{7/2}=N_{9/2}=10^4$ atoms. (a)
  Oscillation frequencies (in units of 1/s) as a function of the
  collision rate $\Gamma_q$ (in units of 1/s).  The circles correspond
  to $T=0.3\,T_F$ while the squares to $T=0.6\,T_F$; 
  filled symbols mark the $m_F=9/2$ species and the empty ones the
  $m_F=7/2$ species. The dash-dotted lines indicate the location of
  $\omega_{HD}$ while the horizontal dashed lines indicate the bare
  trap frequencies. (b) Average damping rate $\bar\gamma$ (in units of
  1/s) as a function of $\Gamma_q$.}
\label{fig_shif_2comp}
\end{figure}

The equilibrium fermionic density profiles for $2\times 10^4$
particles are first created with the traps centred at
$\bi{r}=(0,\,0,\,1)\,a_{ho}$.  At $t=0$ we let the confinements return
to the position $\rv = 0$ and hence the atoms start oscillating.  In
the collisionless regime the natural oscillation frequencies of the two
species are different and correspond to trap frequencies
renormalized by mean-field interactions, whereas the hydrodynamic regime is
characterized by a common oscillation frequency of the two clouds,
whose motions have been locked
by frequent collisions between their particles. 
In this limit {\it classical} kinetic theory
predicts that the common frequency is 
$\omega_{HD}=\sqrt{(\omega_{9/2}^2 + \omega_{7/2}^2)/2}$, which in our case
gives $\omega_{HD} \simeq 117.3\, {\rm s}^{-1}$. This is in agreement
with our numerical results as is shown in
Fig. \ref{fig_shif_2comp}(a), where the in-phase dipolar frequencies
are plotted as functions of the collision rate $\Gamma_q$.

In evaluating the oscillation frequencies and damping rates we have
assumed that the dynamics is well described by an exponentially damped
harmonic motion and fitted each centre-of-mass position $z_j(t)$ with
a function $A_j\cos(\omega_jt + \phi_j) \exp(-\gamma_j t)$. We
calculate the average damping rate as $\bar \gamma =
(\gamma_{9/2}+\gamma_{7/2})/2$ and estimate the uncertainty of the
result by analyzing the dependence of the fit parameters on the
fitting range. This uncertainty is reported in the figures as error
bars.  The intermediate regime is characterized by a strong
damping and larger uncertainties in the single-mode frequencies, which
destroy the coherence of the oscillations.

\begin{figure}
\centering 
\includegraphics[width=0.8\linewidth]{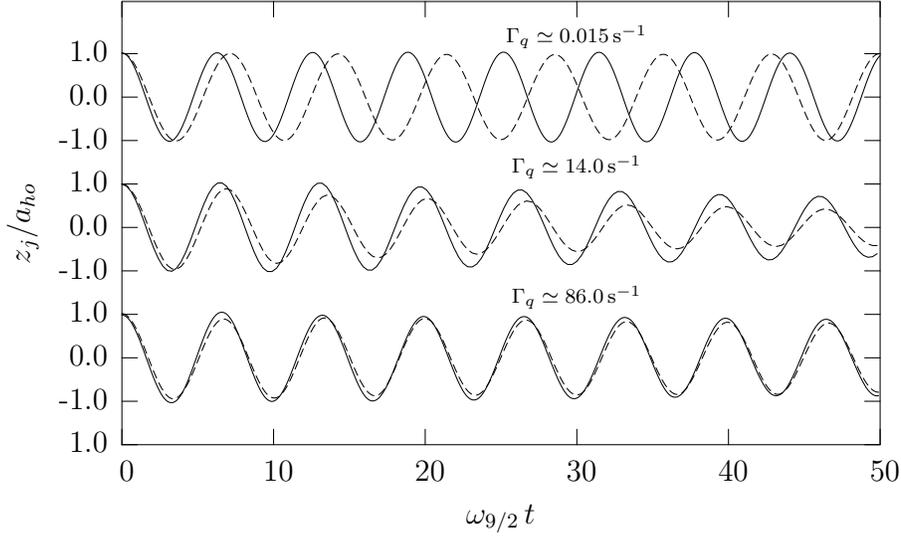}
\caption{\label{fig_z_normal}Centre-of-mass position $z_j(t)$ (in units
  of the oscillator length $a_{ho}=\sqrt{\hbar/m\omega_{9/2}}$) as
  functions of $\omega_{9/2}t$ for a mixture of $^{40}$K atoms at
  $T=0.3T_F$. From top to bottom, the indicated values of the
  collisional rate correspond to $a=150 a_0$ (collisionless regime),
  $a = 5.0\times10^3 a_0$ (intermediate regime), and $a = 1.5\times10^4
  a_0$ (hydrodynamic regime), in units of the Bohr radius $a_0$. }
\end{figure}

As an illustration of the various dynamical regimes we show in
Fig.~\ref{fig_z_normal} the evolution of the centre of mass of the two
species for three values of $\Gamma_q$ at $T=0.3\,T_F$. For
$2\pi\Gamma_q\omega_j^{-1} \ll 1$ the two clouds oscillate
independently without appreciable damping, whereas as
$2\pi\Gamma_q\omega_j^{-1}$ approaches unity the two centres of mass
start moving together but still incoherently and therefore their
oscillations are damped. At higher collisionality
($2\pi\Gamma_q\omega_j^{-1} \gg 1$) the two species are locked and
oscillate at the same frequency without damping.

\begin{figure}
\centering
\includegraphics[width=0.8\linewidth]{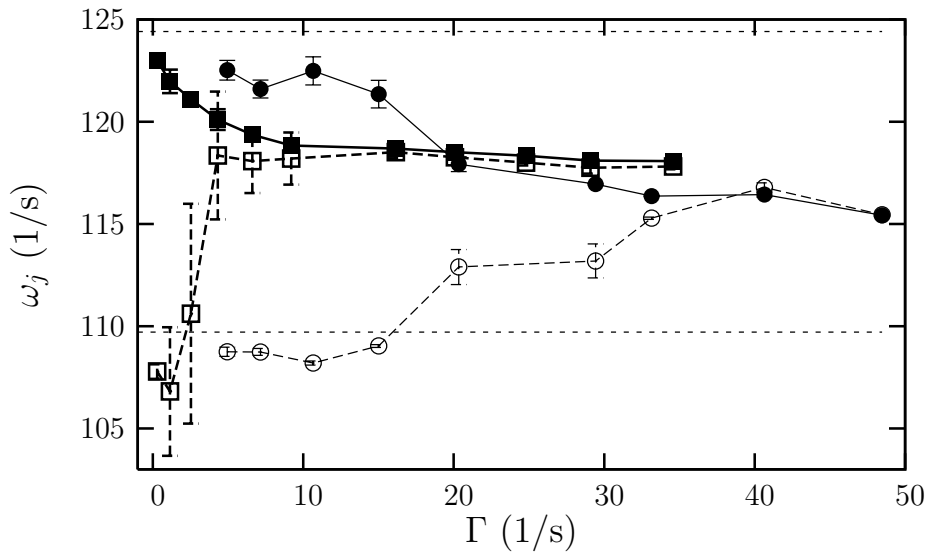}
\caption{\label{fig_comp}Comparison of the oscillation frequencies
  $\omega_j$ (in units of 1/s) obtained with the present 3D approach
  (squares) with those of an effective 2D system (circles) for a
  $^{40}$K mixture with $N_{9/2}=N_{7/2}=100$ particles at $T=0.3\,T_F$.
  The empty symbols correspond to the $m_F=7/2$ species and the filled
  ones to the $m_F=9/2$ one.}
\end{figure}

From Figures \ref{fig_shif_2comp}(a) and \ref{fig_shif_2comp}(b) it is
clear that the transition towards the hydrodynamic regime occurs for
lower values of $\Gamma_q$ at $T=0.3\,T_F$ than at $T=0.6\,T_F$.  This
phenomenon can be attributed to collisions which involve particles in
a narrower energy range around the Fermi level as the temperature
decreases~\cite{Toschi2003a}.  However, the results presented here are
only in qualitative agreement with those obtained in
Refs.~\cite{Toschi2003a}.  In fact, the 2D approach overestimates the
collision rate and the damping rate.  This is shown in
Fig.~\ref{fig_comp} where the oscillation frequencies are compared
with those obtained by using the 2D code.

\subsection{Kicked oscillations} 

We next excite the dipolar mode with a different initial
configuration, by taking the trap centre at $(0,\,0,\,0)$ for the
$m_F=7/2$ component and at $(0,\,0,\,1)\,a_{ho}$ for the $m_F=9/2$
component. The confinement for the $m_F=9/2$ component is then shifted
to $z=0$ and only its atoms start oscillating.  If the two clouds did
not interact, the $m_F=9/2$ component would oscillate at its bare
frequency and the centre-of-mass of the other species would not move.
Collisionality drives both clouds into oscillation.

\begin{figure}
\subfigure[]{\includegraphics[width=0.5\linewidth]{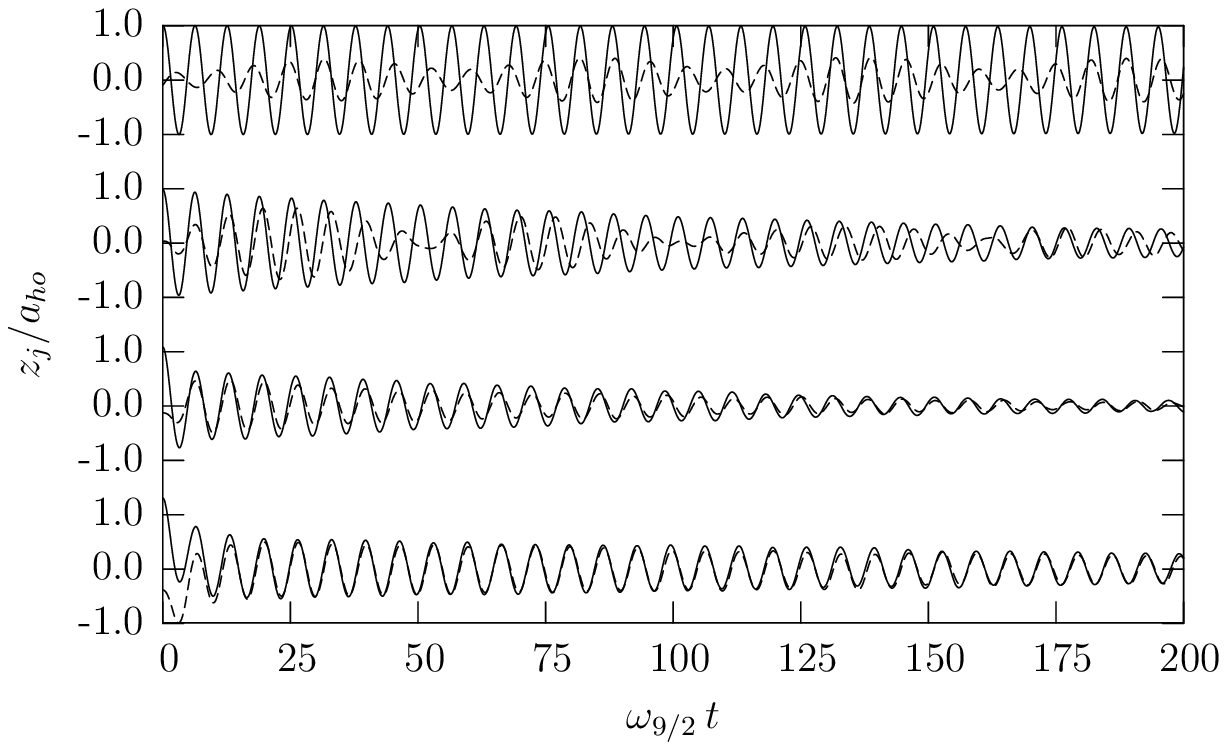}}
\subfigure[]{\includegraphics[width=0.5\linewidth]{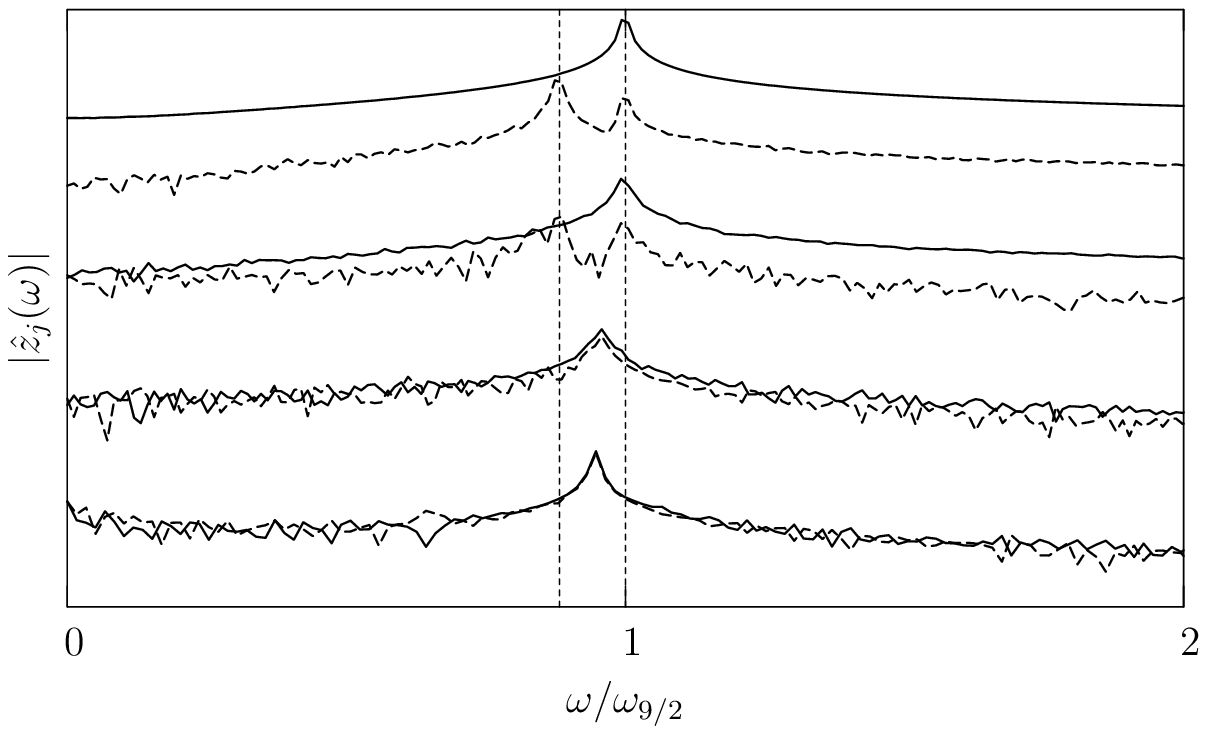}}
\caption{Kicked oscillations of the $m_F=9/2$ component (full lines)
  and the $m_F=7/2$ component (dashed lines) of a $^{40}$K mixture at
  $T = 0.3 T_F$ with $N=2\times 10^4$ atoms.  (a) Centre-of-mass
  positions $z_j(t)$ (in units of the oscillator length $a_{ho}$) as a
  function of $\omega_{9/2}\,t$: from top to bottom the data
  correspond to $a=150\,a_0$ (collisionless regime), $a=1.5\times
  10^3\,a_0$ (intermediate regime), $a=5.0\times 10^3\,a_0$
  (intermediate regime approaching collisional), and $a=1.5\times
  10^4\,a_0$ (hydrodynamic regime). The amplitude of $z_{7/2}(t)$ has
  been rescaled by a factor 10 and 5 in the top and second row,
  respectively. (b) Corresponding Fourier transform $\left|
  \hat{z}_j(\omega)\right|$ (in arbitrary units and log scale) as a
  function of the frequency $\omega$ (in units of $\omega_{9/2}$). The
  vertical dashed lines indicate the location of the bare trap
  frequencies.}
\label{fig_shift_1comp}
\end{figure}

In Fig.~\ref{fig_shift_1comp}(a) we show the axial centre-of-mass
position for the two clouds as a function of time for four values of
the collision rate $\Gamma_q$.  At variance from what was found in the
in-phase case, a single-mode model cannot describe the oscillations
since even in the collisionless region the $m_F=7/2$ motion has a
strong frequency component induced by the $m_F=9/2$ motion.  This can
be seen from the corresponding Fourier transform $\left| \hat
z_j(\omega)\right|$ of the centre-of-mass positions in the various
regimes, which are shown in Fig.\ \ref{fig_shift_1comp}(b).  In the
case of small collisionality (the two top curves in Fig.\
\ref{fig_shift_1comp}) the $m_F=9/2$ component acts as a drive on the
$m_F=7/2$ component but is not strongly affected by collisions.

The corresponding $z_j(t)$ trajectories are drawn in
Fig.~\ref{fig_shift_1comp}(a). For low collisionality (top and second
row in Fig.~\ref{fig_shift_1comp}), a beating between the two peaks in
the $m_F=7/2$ spectrum is visible in the $z_{7/2}(t)$ signal.  At
higher collisionality the two clouds get locked and oscillate at a
single frequency, which is intermediate between the two trap
frequencies (see also the spectrum in Fig.\ref{fig_shift_1comp}(b))).

\section{\label{sec4} Summary and concluding remarks}

We have studied the transition from the collisionless to the
hydrodynamic regime in a $^{40}$K mixture at temperatures where the
effects of Pauli blocking are noticeable.  The Vlasov-Landau equations
of two interacting fermionic fluids are solved numerically by means of
a fully three-dimensional particle-dynamics approach incorporating
mean-field and collisional interactions.  The implementation of a
\textit{locally-adaptive importance sampling} technique during the
collisional step has allowed us to obtain a method which is very
efficient at low temperature, for realistic number of particles and in
a wide range of collisional strength.  By exploiting this code we
investigate the dipolar collective modes and analyze the oscillation
frequencies and damping rates of the system for various initial
configurations. We find that the transition to hydrodynamics is
shifted to lower values of the collision rate as temperature
decreases.

As a future application, this procedure shall permit us to perform
numerical experiments on cooling and expansion dynamics. These studies
will help characterize the mixture in the intermediate regime where
neither purely collisionless nor collisional approaches are
appropriate. In addition the extension to highly interacting systems
in the so-called \textit{unitary} limit of collisions \cite{Gehm2003b}
shall provide more physical insight into recent experiments on
\textit{attractive} mixtures of fermionic atoms
\cite{OHara2002b}. Efforts to pursue these further studies are
currently underway and will be reported elsewhere.

\ack This work has been partially supported by INFM under the
PRA-Photonmatter Programme.

\section*{References}


\end{document}